# Convergence of IT and SCADA: Associated Security Threats and Vulnerabilities

**Michael Smurthwaite and Maumita Bhattacharya***

School of Computing & Mathematics, Charles Sturt University, Australia

*Corresponding author e-mail: mbhattacharya@csu.edu.au

**Abstract.** As many industries shift towards centralised controlled information systems for monitoring and control, more importance is being placed upon technologies such as Supervisory Control and Data Acquisitions industrial systems (SCADA). This focus on integration and interoperability presents numerous challenges for security personnel and organisational management alike. It becomes paramount therefore to reciprocate this new direction within an organisation with adequate plans and frameworks that ensure protection and security of its SCADA architecture. A clear understanding of the relevant threats and vulnerabilities is critical for adopting/developing appropriate policy and frameworks. To this end, in this research we identify and analyse relevant SCADA security threats and vulnerabilities and present a simple scheme to classify them for better understanding.

## 1. Introduction

It is recognised globally that the convergence between Information Technology (IT) and Operations Technology (OT) or Industrial Control Systems (ICS) environment(s), is blurring both functional and operational boundaries within organisations, be this through technological or organisational change. This current day tendency on interoperability and integration between networks often is intended to provide organisational capabilities for real-time optimization of business processes, attempting to bolster corporate efficiency, competitiveness and organisational productivity. A focal point to this is the consolidation of organisational data, so as to provide integrated enterprise information management ensuring that the best possible decision can be made from the right information, in the right format, to the right person, at the right time. This presents a fundamental dilemma between resources that are and have been divergent since inception; as such the criticality of the OT environment cannot be underestimated to an organisation, in particular the protection of embedded SCADA (Supervisory Control and Data Acquisition) software systems which, due to intricacies, provide a challenging situation for management and security [5, 11, 13]. This convergence dictates the need for a more thorough and formulated approach in the management of inherent risks. The premise of our current research is to investigate the SCADA security threats and vulnerabilities and introduce a simple scheme to classify them for clearer understanding. In effect, the point of classification is to inform the process of adoption/development of appropriate policy and framework for risk management. At an operational level, it can be also used to properly communicate to relevant staff and business leaders where there is an area of risk that requires attention.

The rest of the paper is organised as follows: Section 2 briefly explains the inherent generational weaknesses of SCADA systems. An analysis of SCADA security threats and vulnerabilities are presented in Section 3. In Section 4, we introduce the scheme for classification of SCADA security threats and vulnerabilities and finally, concluding remarks are presented in Section 5.

## 2. SCADA systems: The inherent weaknesses

The core purpose of the OT or ICS environment is to deliver a setting for collective real-time control, sensing and monitoring capabilities as part of the service operations processes across an organisation. The ICS environments typically can be considered fault tolerant and tend to rely on layered configuration, whereby data traverses a series of layers between devices and sensors in the field and those that are used to control those devices [2, 3]. The strategic importance of ICS or SCADA related systems, and how these contribute to a functioning society cannot be emphasised enough. The spectrum of applications and functions that are delivered across the globe, can in essence be considered at the heart of many things, we take for granted; for example: electricity, water and gas transmission and distribution to various public and private industrial process monitoring [4].

For the *first* and *second* generation of SCADA systems, implementation focal point was purely from an operational perspective, failing to account for security aspects during the process. Therefore, by leveraging the fundamental nuances associated within the systems, including proprietary software/protocols or dedicated communication paths to devices, in combination with minimal network technological adoption and isolation or the 'air gap' re-affirmed an organisation's belief that there system was secure [2, 15]. Security through obscurity can also be considered a primary aspect of both generations of systems. With regards to current day systems, the distinction between physical and logical independence from corporate networks can be considered the fundamental flashpoint. With the reliance on open source technologies and common protocols, not to mention the focus on organisational interoperability, the SCADA or ICS environment now contains much of the same flexibility as an IT environment, as well as the potential drawbacks associated with increased interconnectivity and potential vulnerabilities [1, 5, 6]. Systemically this is at the heart of many of the security concerns facing management and security professionals. These environments differ architecturally and also the management and possible repercussions of a successful attack are radically different. Table 1 summarises the key generational weaknesses of SCADA systems.

**Table 1.** Generational weaknesses of SCADA systems

| SCADA Generation | Time Period | Weaknesses |
|---|---|---|
| **1st Generation** | 1960s | The inherent weakness of this generations include:<br>• Co-location of server and users.<br>• Built on Mainframe Technologies; if mainframe becomes unavailable the so does the SCADA abilities.<br>• Reliance on proprietary software and protocols on dedicated paths to devices.<br>• 'Security through Obscurity', and protection through isolation or 'Air Gap'. |
| **2nd Generation** | 1980s | Incorporating the many of the weaknesses from 1st Generation, further weaknesses include:<br>• Limited geographical extent.<br>• Technologically reliant on evolving network technologies.<br>• Communication protocols could be easily read and understood. |
| **Current Day** | >2000 to present | These include:<br>• Built on open standards and architecture, moving away from proprietary standards.<br>• Fundamental reliance on network technologies and connectivity.<br>• Interconnection and interoperability between corporate and SCADA networks.<br>• Utilisation of web technologies opens organisation to cyber-attacks. |

## 3. Security threats and vulnerabilities: An analysis

Not unlike conventional ICT related systems, SCADA systems face many security threats and vulnerabilities. This section attempts to identify and analyse them, while providing a simple classification mechanism in Section 4, which avoids any ambiguity. Similarly to [2], fundamental security differences will be highlighted, alluding to the elementary security aspects that compared to the IT environment are often accepted risks due to the importance of that component or device. Between out-dated Operating Systems, inability to arrange downtime, compatibility or concerns over the lasting impact of an applied patch, the ICS environment does offer a peculiar set of security risks for professionals to adapt too.

Elaborated within NIST Risk management guide, a threat source can be considered the concurrence of an actor's interaction and motivation. Establishing a taxonomy of threat source, they break this down into four overarching categories: Adversarial, Accidental, Structural and Environmental [7, 12]. This definition is centred on actions taken by an actor or threat agent, be this with malicious intent or not.

Many of the threat sources facing SCADA systems are built upon the generational weaknesses elaborated in the previous section. These foundations provide substantial opportunity for attack vectors from multiple areas. However despite these, outside generational weaknesses, supplementary threat sources continue to be discovered and built upon.

### 3.1. Common threats and vulnerabilities

The most glaring common aspect that has led to numerous threat sources, which is rarely furthered in discussions, can be considered the divide that exists between Information Technology and SCADA related teams, be this political, cultural or technical. Historically separate as discussed, the emergence of organisational and business drivers forcing this cohesion, has often led to disparate and an environment that is segmented with no demarcation between operational duties. This incongruence is obvious with the differing mentality with regards to SCADA systems, in particular as seen by SCADA operators, the threat to safety and continuity of the process, far outweighs the application of IT security technologies [10].

The increasing interconnectivity and integration of SCADA systems has also amplified the exposure of the network to a greater number of access points, this not only provides the need for greater understanding but compounds the complexity of the environment with a greater number of subsystems that have interconnections between each other [8, 9]. This further makes redundant the older security beliefs of 'air gap' or security through obscurity, and disconnection from internet.

Technologically, the threat sources faced via the use of common computing technologies makes SCADA systems susceptible to many of their inherent weaknesses. It is not uncommon for legacy systems to be operating side by side with modern solutions, with both of these rarely under patch management. This coupled with the implementation of these systems often being based upon a 'set and forget' mentality, whereby factory standard settings or default passwords are left as part of the installation, has resulted in many SCADA systems operating a variety of hardware and software configurations, some of which are no longer supported but are critical for operational functionality.

This is further hindered operationally due to the inability to make changes which would breach contractual obligations or vendor warranty specification. In effect, due to the inability for vendors to remain and guarantee a technology, current upgrade path opens up many situations for manipulation or exploitation by multiple threat sources. The ability to disable security settings for ease of installation and usability also should be considered as a configuration option that is lightly used.

This leads to a failure by most SCADA system owners to actively and continuously apply both system wide and system specific configuration management. This critical aspect of security affords an organisation to actively monitor and track configurations, thus ensuring that any authorised changes are captured and threat sources that have made these changes, are understood. The lack of security and configuration baselines for many of the SCADA systems that exist, really highlights the inability to detect abnormal activity.

Other common attack vectors that should be considered, as is the case with any form of IT Technology include: (i) Online threats including malware, phishing scams etc. (ii) Internal threats. (iii) External threats, such as, industrial espionage or 'hacktivism'. (iv) Mobile storage mediums and mobile devices.

## 4. Threat and vulnerability classification: A simple scheme

While classifying SCADA security threats and vulnerabilities, it is best to first assess the risk by identify the following:

- Impact (What event would impact the business and to what maximum extent).
- Actor (Who would exploit this, and for what reason or who could accidentally become the trigger for the event).
- Source or Attack Vector (Where would the exploitation originate).
- Likelihood (Combines types of vulnerability, such as, physical/logical/technical, and any effective mitigation in place. Considering them, what is the likelihood that the event would occur within the current environment).

To classify threats and vulnerabilities is therefore a process of using the risk as assessed above and then reporting it in the context of both the organisation's risk appetite and a description of the impact to business operations. To do this effectively one must develop a set of terms that is understood by the business. One of the key intents of classification is to properly communicate to relevant staff and business leaders, where there is an area of risk that requires attention. For example: You may determine, a technical vulnerability exists within the configuration of a system. Then you assess the maximum impact of exploiting this vulnerability and the extended failure of the Network Management System. You may determine, there is only one attack vector that could exploit that vulnerability and it could only happen from inside the network environment. Next, you determine the likelihood of occurrence, based on historical industry data. Using these information, you consult the enterprise risk register, to assess the risk as HIGH. Now you need to classify the risk in terms of 'business context'. For example: 'There is a HIGH risk causing disruption for scheduled network maintenance work'.

**Table 2.** Classification of threat/vulnerability sources based on the proposed scheme

| Dimension | Physical | Logical |
|---|---|---|
| **People** | Incorporates the human related threats to the Physical Infrastructure, including theft and damage. | From the logical perspective, SCADA security threats in relation to the people category; an example of such is insufficient access control process in place to verify the user issued command. |
| **Process** | Combines the processes in place, for installation, maintenance and decommissioning of devices. | Formulation of the processes. |
| **Network/Integration** | Covers the Network and Integration at a physical level, for example, Reliance on network technologies and public communication, VPN connectivity. | Fundamentally can be considered one of the biggest threats of a SCADA system, from threats associated to protocols as well as data simplicity. Segregation can also be considered a particularly relevant component. |
| **Technical/Technological** | Comprising of physical technological threats; both with historical and future perspectives; for example, ageing infrastructure could be a historical technical threat. | Covers the technological threats at logical layers; for example, out-dated or unpatched operating systems or software as well as the adoption of internet based technologies. |

This interconnection and focus on interoperability has brought many of the weakness associated with the corporate ICT environments without the ability to utilise consistently the solutions that attempt to mitigate these, due to functional differences that exist. Moving from isolation to consistent connections between organisational systems including MIS (Management Information Systems), CIS (Customer Information Systems), DMS (Distribution Management Systems) and EMS (Energy Management Systems), to name a few, and the utilisation of communication technologies and ICT technologies have further enabled the potential risks of exploits being applicable.

Considering the above factors, we introduce the following four dimensions, in order to classify various components of threats and vulnerabilities, in the context of SCADA and IT convergence, in a simple, yet adequate manner: (i) *People*, (ii) *Process*, (iii) *Network/Integration* and (iv) *Technical/Technological*. As an example, Table 2 presents a classification of the *sources* of threat and vulnerability, based on our proposed scheme. Similar classification is possible for types of threats and vulnerabilities, potential control mechanisms and so on. The strength of our proposed scheme is its simplicity.

In this context, it may be noted that our proposed classification scheme is partly influenced by [2], where, Cheminod et al. characterised and tabularized system characteristics, typical security practices and countermeasures, utilised between corporate and ICS environments, highlighting the fundamental discrepancies that exist due to different and overarching intended purposes and design of each environment.

*4.1. Moving forward*

The distributed and networked nature of many of the current SCADA systems opens a number of possibilities for intrusion through varying communication protocols. These connections have the potential to provide greater numbers of access points while increasing system and environment complexities [11, 13]. Moving forward, any discussion around the current day SCADA threats and vulnerabilities [14, 15] should also include, security issues associated with the use of ICT infrastructure and the previous security focus within older generations of SCADA systems. Some examples are as follows:

- Openness to electronic attack through the utilisation of and reliance on network technologies. (The widening of distribution networks has forced a reliance on public telecommunication links for interconnectivity instead of the previous separate links designated for devices.)
- Interconnection between corporate and SCADA networks. (These linkages to corporate information systems, allow for the potential undesirable access to the SCADA network.)
- The diversity of vendors, in particular the use of common standards. (Although this has provided advantages with regards to integration of devices and hardware, the open standards and protocols expose SCADA systems to manipulation of these protocols and technologies.)
- The ageing of equipment and the inability to interact with modern day technologies.
- Data simplicity and real-time processing. (The very nature of the system, due to exposure, is now one of SCADA's greatest vulnerabilities. Also included is the failure to provide confidentiality when transferring data.)
- The large scale lack of authentication that exists across many SCADA systems, from the use of default installation passwords to no passwords at all. (The ability to accurately authenticate the user's identity when performing a function is not always the strongest.)

**5. Conclusions**

Securing SCADA environments provides a number of challenges to security practitioners and managers alike. The largest deterministic value for this challenge is the inherent nature and disparity that commonly exists between SCADA and traditional IT systems. In this paper, we have explored and analysed some of the key SCADA security threats and vulnerabilities and introduced a simple scheme for classifying these threats and vulnerabilities to assist adoption of appropriate mitigation strategies.

Due to the utilisation and application of SCADA or IACS technologies, across a variety of industries, not to mention distribution, the difficulty in establishing and utilising an overarching holistic security framework should be highlighted. Internationally, there are a number of agencies, which provide advice and recommendations on good practices. The common theme for this seems to be the focus on stakeholder involvement, engaging stakeholders in open discussion on the protection of their system, in an attempt to address the major concerns facing operators within this area. Industry best practices provide assistance to organisations. However, more often the case is that organisations are left to pick and choose what they consider applicable to themselves. Fundamentally there is no fault with this process. However, if there were more applicable industry specific methodologies, in combination with overarching holistic processes, we would find a more integrated security practice that should address many of the underlying flaws. The growing awareness has improved security somewhat, over the past decade. However without greater engagement and interaction, not just at any national level, but an international level, it would be difficult to see a mutually beneficial outcome.